\documentclass{mem}
\usepackage{natbib}\usepackage{txfonts}\usepackage{balance}
\usepackage{graphicx}
\idline{1}{1}
\begin{document}
\newcommand\chandra{\textsl{Chandra}}
\newcommand\cyg{Cyg~X-1}
\newcommand\gso{{GSO}}
\newcommand\hetg{{HETG}}
\newcommand\hexte{{HEXTE}}
\newcommand\hxd{{HXD}}
\newcommand\integral{\textsl{INTEGRAL}}
\newcommand\pca{{PCA}}
\newcommand\rxte{\textsl{RXTE}}
\newcommand\suzaku{\textsl{Suzaku}}
\newcommand\xis{{XIS}}
\newcommand\spi{{SPI}}

\title{
The Microquasar Cyg X-1: A Short Review
}

   \subtitle{}

\author{
M. \, A. \, Nowak\inst{1},
J. \, Wilms\inst{2},
M. \, Hanke\inst{2},
K. \, Pottschmidt\inst{3}, \and
S. \, Markoff\inst{4}
          }

  \offprints{M. A. Nowak}

\institute{
Massachusetts Institute of Technology, Kavli Institute for
Astrophysics, Cambridge, MA 02139, USA
\and
Dr.~Karl Remeis-Sternwarte and Erlangen Centre for Astroparticle
Physics, Universit\"at Erlangen-N\"urnberg, Sternwartstr.~7, 96049
Bamberg, Germany
\and
CRESST, UMBC, and NASA Goddard Space Flight Center, Greenbelt, MD
20771
\and
Astronomical Institute ``Anton Pannekoek'' University of Amsterdam\\
\email{mnowak@space.mit.edu}
}

\authorrunning{Nowak }

\titlerunning{The Microquasar Cyg X-1}

\abstract{We review the spectral properties of the black hole
  candidate Cygnus~X-1.  Specifically, we discuss two recent sets of
  multi-satellite observations.  One comprises a 0.5--500\,keV
  spectrum, obtained with \emph{every} flying X-ray satellite at that
  time, that is among the hardest \cyg\ spectra observed to date.  The
  second set is comprised of 0.5--40\,keV \chandra-\hetg\ plus
  \rxte-\pca\ spectra from a radio-quiet, spectrally soft state.  We
  first discuss the ``messy astrophysics'' often neglected in the
  study of \cyg, i.e., ionized absorption from the wind of the
  secondary and the foreground dust scattering halo. We then discuss
  components common to both state extremes: a low temperature
  accretion disk, and a relativistically broadened Fe line and
  reflection.  Hard state spectral models indicate that the disk inner
  edge does \emph{not} extend beyond $\gtrsim 40\,GM/c^2$, and may
  even approach as close as $\approx 6\,GM/c^2$. The soft state
  exhibits a much more prominent disk component; however, its very low
  normalization plausibly indicates a spinning black hole in the
  \cyg\ system.  \keywords{accretion, accretion disks -- black hole
    physics -- X-rays:binaries \vspace{-2mm}}} \maketitle{}

\section{Introduction}

After the initial discovery \citep{bowyer:65a} of the black hole
candidate \cyg, \citet{tananbaum:72a} noted that, at least in the
2--10\,keV band, it exhibited two distinct spectral states: a bright,
spectrally soft state, and a fainter, spectrally hard state.
Subsequent broad-band studies further elucidated this dichotomy.  The
soft state spectrum peaks between 1--2\,keV, has a weak hard tail with
photon spectral index $\Gamma > 2.1$, and is radio quiet.  The hard
state spectrum is well-described by an exponentially cutoff (folding
energy 125--250\,keV) broken powerlaw with 2-10\,keV photon index
$\Gamma < 2.1$ and a harder photon index at $\gtrsim 10$\,keV, weak
soft excess, and it is radio loud.  \cyg\ spends the majority of its
time in its hard state.  A range of such spectra (and further
references) can be found in the work of \citet{wilms:06a}, which
describes a multi-year radio/\rxte\ monitoring campaign of \cyg.

\begin{figure}[]
\resizebox{\hsize}{!}{\includegraphics[clip=true]{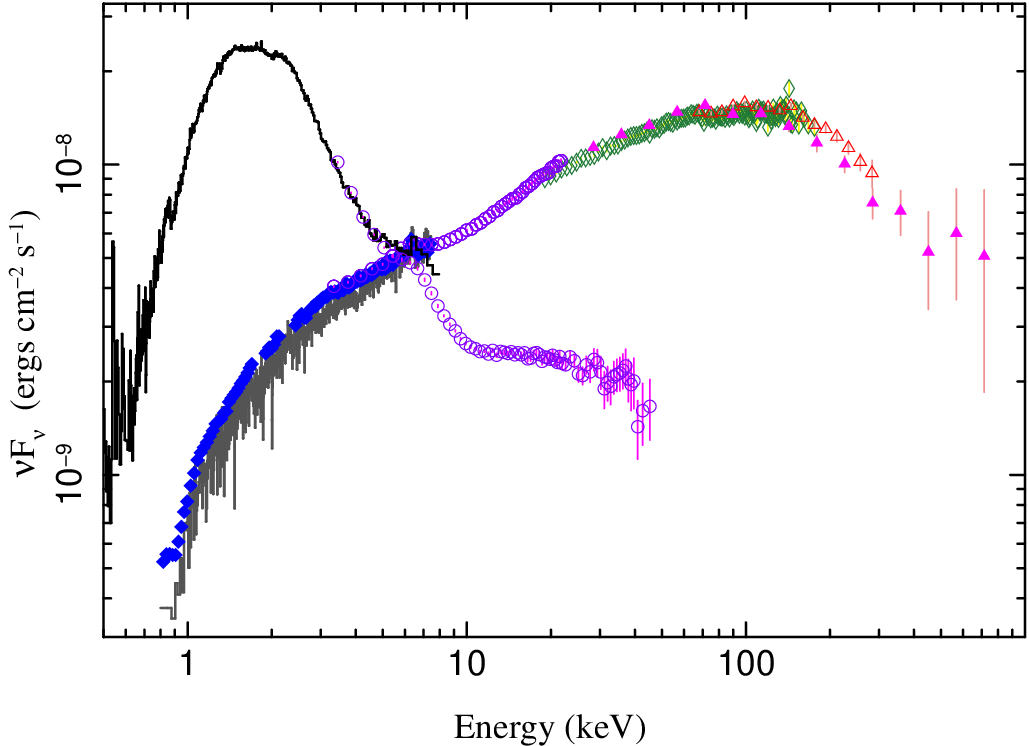}}
\caption{ \footnotesize \cyg\ in its two most extreme spectral states:
  the hard state dominated by a cutoff powerlaw spectrum extending to
  $> 100$\,keV, and the soft state dominated by a thermal spectrum
  peaking between 1--2\,keV.  The above hard state is from a
  multi-wavelength campaign wherein \cyg\ was observed by \emph{every}
  flying X-ray satellite.  Here we display \chandra-\hetg\ spectra
  (histogram), \suzaku-\xis\ (solid diamonds) and -\gso\ (hollow
  triangles) spectra, \rxte-\pca\ (circles) and -\hexte\ (hollow
  diamonds) spectra, and \integral-\spi\ (solid triangles) spectra
  (see \protect{\citealt{nowak:11a}}).  The soft state spectra are
  from a simultaneous \chandra-\hetg\ and \rxte\ campaign conducted in
  January 2011. The above spectra are presented without reference to
  any underlying spectral model, and are unfolded using only the
  spectral response of the detectors (see
  \protect{\citealt{nowak:05a}}).}
\label{fig:states}
\end{figure}

Two extreme examples of the states are presented in
Fig.~\ref{fig:states}.  This figure shows April 2008 observations of a
spectrally hard state, covering the bandpass of 0.5--500\,keV.  It is
among the hardest spectra seen from \cyg; a description of these
spectra can be found in \citet{nowak:11a}.  Also shown are spectra
from simultaneous \chandra-\hetg/\rxte-\pca\ observations, covering
the 0.5--40\,keV range of a spectrally soft state.  In terms of
\emph{observed} flux, the 0.1--50\,keV flux of the soft state is $3.9
\times 10^{-8}\,{\rm erg\,cm^{-2}\,s^{-1}}$, while the \emph{observed}
0.5--300\,keV flux of the hard state is $4.9 \times 10^{-8}\,{\rm
  erg\,cm^{-2}\,s^{-1}}$.  However, making plausible corrections for
absorption and extrapolating the spectra to determine bolometric
luminosities, this is at 2.2\%\,${\rm L}_{\rm Edd}$ and this hard
state is at 1.6\%\,${\rm L}_{\rm Edd}$.  (The Eddington luminosity,
${\rm L_{\rm Edd}}$, used here assumes a distance of 1.86\,kpc and a
mass of 15\,${\rm M}_\odot$; \citealt{reid:11a,orosz:11a}.)  That is,
whereas we directly observe most of the flux in the hard state, nearly
2/3 of the soft state flux is \emph{unobserved} due to both absorption
and bandpass limitations.

\begin{figure}[]
\resizebox{\hsize}{!}{\includegraphics[clip=true,bb=30 340 550
    770]{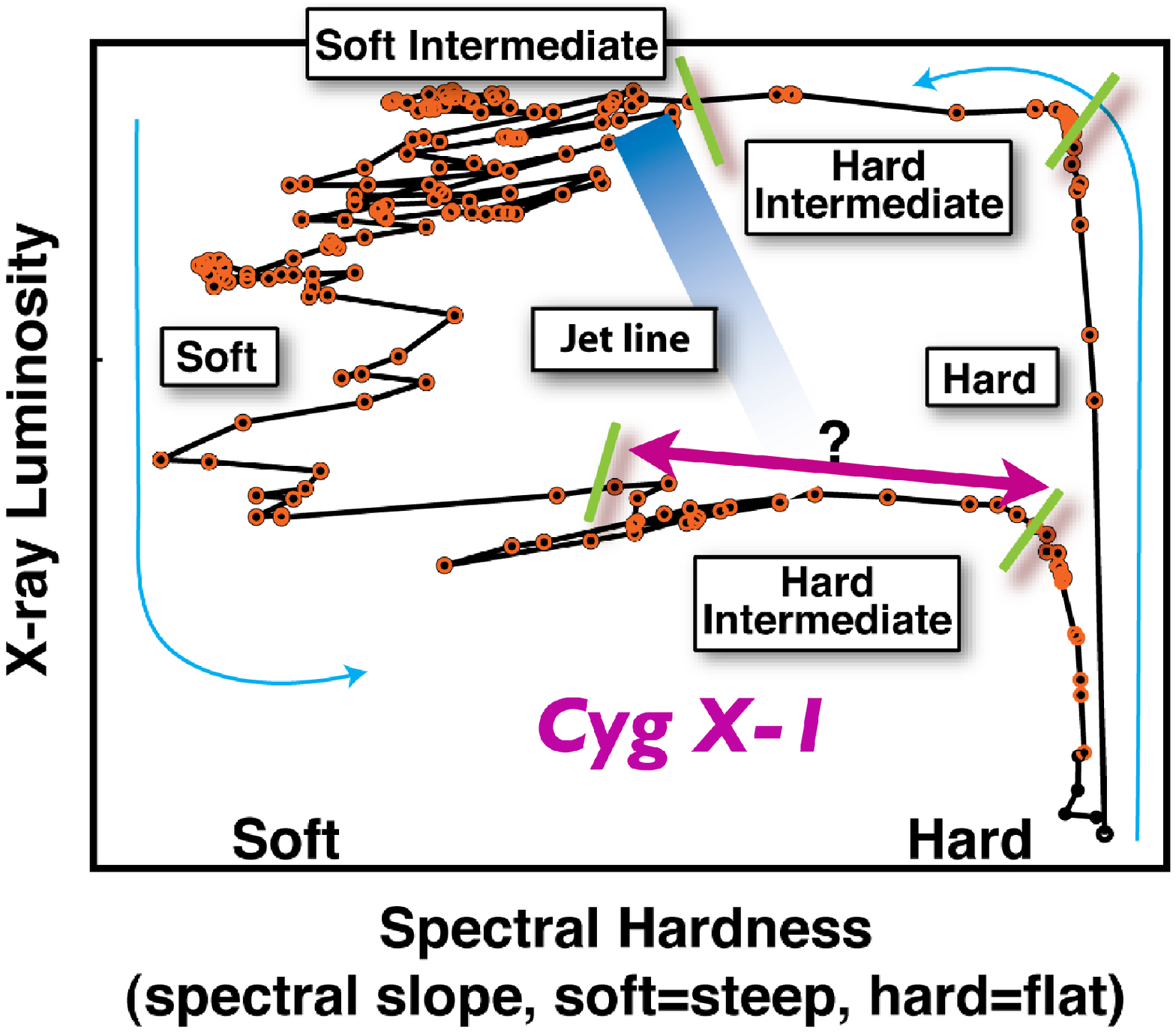}}
\caption{ \footnotesize A typical hardness-intensity diagram for a
  black hole transient outburst (based upon an outburst of GX~339$-$4;
  \protect{\citealt{belloni:05a,homan:05a}}).  Transients begin faint,
  hard, and radio loud; they evolve to brighter states while remaining
  hard; they soften and become radio quiet (often preceded by a radio
  ejection event); they then fade, harden, and become radio loud once
  more.  \cyg, which is persistently emitting in the X-ray band,
  occupies only a small portion of this diagram at fractional
  Eddington luminosities of $\approx 2\%$, near the so-called radio
  loud/radio quiet transition ``jet line''.}
\label{fig:q}
\end{figure}

As has been noted by previous researchers, the range of bolometric
luminosities traversed by \cyg\ spans only a factor of $\approx 3$--4
\citep[][and references therein]{wilms:06a}.  This is somewhat narrow
compared to most black hole transients.  Furthermore, \cyg\ also
traverses a narrower range of colors than many black holes, never
exhibiting a purely disk-dominated spectrum without a hard tail
\citep[e.g., like the simple disk-dominated spectrum of
  4U~1957+11;][]{nowak:08a}.  Black hole transients often follow
color-intensity diagrams as shown in Fig.~\ref{fig:q}, spanning
$\gtrsim 3$ orders of magnitude in luminosity, and wider extremes of
color variations.  \cyg, however, exists on an especially interesting
portion of this `q-diagram' --- moving between the radio quiet/soft
$\Leftrightarrow$ radio loud/hard-intermediate state transition near
$\approx 2$\%\,${\rm L}_{\rm Edd}$.

\begin{figure}[]
\resizebox{\hsize}{!}{\includegraphics[clip=true]{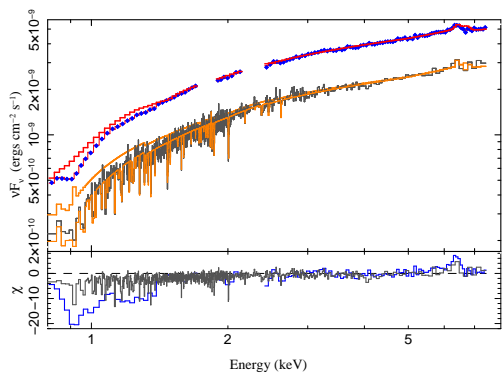}}
\caption{ \footnotesize Simultaneous \suzaku-\xis\ (diamonds) and
  \chandra-\hetg\ (histogram) spectra fit with a simple
  phenomenological model consisting of a disk, powerlaw, a
  relativistically broadened line, and a narrow line, modified by
  ionized absorption (see \protect{\citealt{hanke:08a}} and
  \protect{\citealt{nowak:11a}}). Furthermore, the \hetg\ spectra are
  modified by dust scattering.  One set of models, and all of the
  residuals, are shown with both the ionized absorption and dust
  scattering absent.}\label{fig:messy}
\end{figure}

The hypothesized emission components in the \cyg\ system include an
accretion disk, a Comptonizing corona (with either a thermal, or
hybrid thermal/non-thermal electron distribution; \citealt{coppi:99a})
that upscatters photons from the disk, a radio emitting jet that might
also contribute to the X-ray \citep{markoff:05a}, and relativistically
smeared reflection/fluorescence from the disk.  The important
questions are, what are the relative contributions of these
components, and how do they and the system geometry change between the
two state extremes?

\begin{figure}[ht]
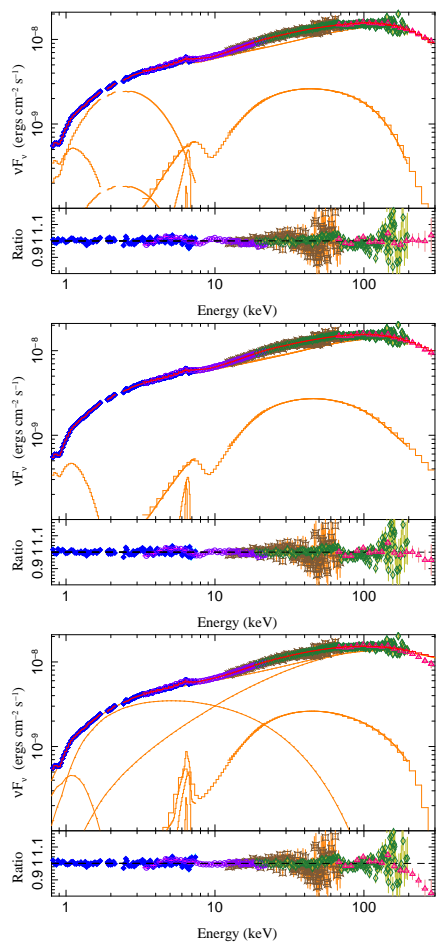

\resizebox{0.86\hsize}{!}{\includegraphics[clip=true]{nowak_2011_01_fig04a.ps}}
\resizebox{0.86\hsize}{!}{\includegraphics[clip=true]{nowak_2011_01_fig04b.ps}}
\resizebox{0.86\hsize}{!}{\includegraphics[clip=true]{nowak_2011_01_fig04c.ps}}
\caption{\footnotesize \cyg\ hard state \suzaku/\rxte\ spectra, fit
  with different models. (Model components shown individually: seed
  photons, Compton spectrum, synchrotron, SSC, disk, reflection, broad
  and narrow lines.)  Top: thermal Comptonization with a high seed
  photon temperature.  Middle: non-thermal Comptonization with a low
  seed photon temperature.  Bottom: jet model dominated by synchrotron
  and SSC emission.  All models also have: a low temperature disk,
  relativistically broadened reflection and Fe fluorescence line, and
  a narrow Fe line.  Disk and broad Fe line parameters are consistent
  with the disk extending very close to the innermost stable circular
  orbit \protect{\citep{nowak:11a}}.}
\label{fig:hard}
\end{figure}

\section{Messy Astrophysics}

Often neglected in models of \cyg\ are two important components that
directly bear upon fits of the low temperature disk component in the
hard state.  First is the fact that \cyg\ has an O-star secondary with
a highly ionized wind that leads to pronounced ionized absorption
\citep{hanke:08a}.  This absorption is spectrally resolved by
\chandra-\hetg\, but not by \suzaku-\xis\ spectra \citep{nowak:11a}.
Second is the fact that foreground dust scattering acts as a loss term
for the 1$^{\prime\prime}$ spatial resolution \chandra\ spectra, but
not for 2$^{\prime}$ spatial resolution \suzaku\ spectra.  For the
latter, the emission scatters back into the line of sight, albeit time
delayed, from an extended dust scattering halo \citep[see][for models
  of the \cyg\ halo, and use of the halo time delay to derive a
  distance consistent with the radio parallax measurements of
  \citealt{reid:11a}]{xiang:11a}.

Both of these components must be included in the fits, as
demonstrated in Fig.~\ref{fig:messy} \citep{nowak:11a}.

\section{Hard State Spectra}

There are a number of questions as to the physical mechanisms
responsible for the hard state spectrum of \cyg. The emission is
typically attributed to a hot corona upscattering photons from an
accretion disk; however, the geometry of this configuration is still
debated.  Does the corona lie central to a truncated outer thin disk
\citep{dove:97b}, or, if the corona is driven outwards by radiative
pressure \citep[e.g.,][] {beloborodov:99a}, could it instead overlay
the inner disk?  Can an optically thick, geometrically thin disk
extend inward nearly to the innermost stable circular orbit
\citep{miller:06b}?  Is the hard state corona comprised primarily of
electrons with a thermal population \citep{poutanen:09a}, or can it
have a substantial contribution from a non-thermal electron population
\citep{ibragimov:05a}?  Alternatively could the X-rays be comprised of
a combination of direct synchrotron and synchrotron self-Compton (SSC)
emission from a jet, in addition to coronal emission
\citep{markoff:05a,maitra:09a}?

In Fig.~\ref{fig:hard} we present a number of these possibilities,
which fit nearly equally well: a thermal corona, a non-thermal
corona, and a jet plus corona model \citep{nowak:11a}.  All three
models, however, have aspects in common.  There is spectral hardening above
10\,keV that is \emph{partly}, but \emph{not solely}, attributable to
reflection.  The spectra require a low temperature disk (peak
temperature 150--250\,eV) with normalization sufficiently low to
indicate a disk inner radius $\lesssim 40\,GM/c^2$ (or even consistent
with $6\,GM/c^2$).  In addition to the narrow Fe line there is a broad
Fe line, also with parameters indicating a disk inner radius of
$6$--$40\,GM/c^2$.

\section{Soft State Spectra}

\begin{figure}[]
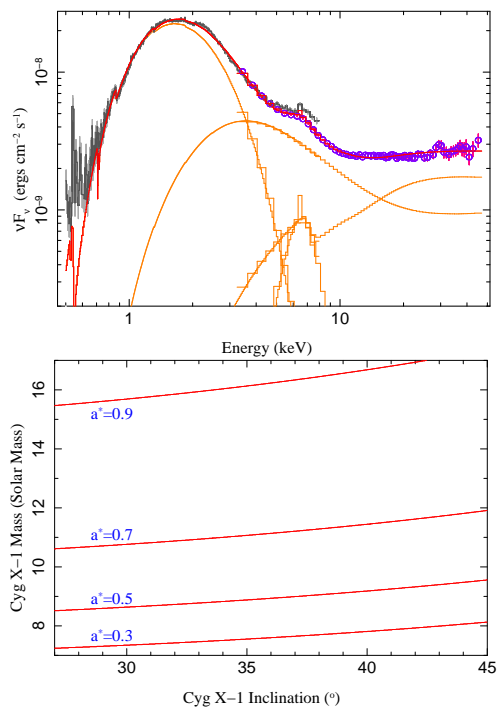

\resizebox{0.97\hsize}{!}{\includegraphics[clip=true]{nowak_2011_01_fig05a.ps}}
\resizebox{\hsize}{!}{\includegraphics[clip=true]{nowak_2011_01_fig05b.ps}}
\vspace{-2mm}
\caption{ \footnotesize Top: The soft state \chandra-\hetg\ and
  \rxte-\pca\ spectra of \cyg\ fit with a non-thermal Comptonization
  model with a dominant, high temperature seed photon spectrum from a
  disk.  Individual model components are shown, including the required
  relativistically smeared reflection/Fe fluorescence.  Bottom: Based
  upon the fitted disk normalization, a simple estimate of the implied
  \cyg\ black hole spin in dimensionless angular momentum units
  vs. black hole mass and system inclination.  We assume a disk
  temperature color correction of 1.7 and a 1.86\,kpc distance.)}
\vspace{-4mm}
\label{fig:soft}
\end{figure}

Soft state spectra, from a January 2011 \chandra-\hetg\ plus
\rxte-\pca\ observation, are shown in Fig.~\ref{fig:soft}.  There is
an absence of radio emission, implying a quenched jet component.
(Ionized absorption is also mostly absent.) Otherwise, the soft state
spectrum contains the same components as the hard state, albeit in
different proportions.  Here the disk is of higher temperature (peak
$kT\approx 470$\,eV), and is clearly dominant.  A hard tail remains,
which we here model with purely non-thermal Comptonization.  A
relativistically broadened line and smeared reflection are also
indicated.

As has been discussed by \citet{gou:11a}, the normalization of the
disk component (here, the seed photons for Comptonization) is rather
small, given recent distance and mass estimates
\citep{reid:11a,orosz:11a}, \emph{and} assuming a color-correction
factor of $1.7$ for the disk spectrum. \citet{gou:11a}, using a fairly
narrow range of masses around $\approx 15\,{\rm M_\odot}$ and
inclinations near $27^\circ$, claim a near maximally spinning black
hole.  Taking a wider range of inclinations and masses, however, still
implies some degree of spin to achieve the low disk normalization,
unless the black hole is at the lowest end of mass ranges from the
literature.

If \cyg\ is rapidly spinning, then it is possible that the hard state
can have an inner disk radius of $\approx 10\,GM/c^2$ (consistent with
our hard state spectral fits), and yet have an inner disk radius a
factor of several times smaller in its spectrally soft state.

\begin{acknowledgements}
We would like to thank the \rxte, \suzaku, and \chandra\ schedulers for
making the presented simultaneous observations possible.  This work
was supported by \suzaku\ and \chandra\ guest observer grants,
NNX07AF71G, NNX08AE23G, NNX08AZ66G, GO8-9036X, G01-12065X, as well as
NASA Grant SV3-73016. J. Wilms and M. Hanke would like to acknowledge
the support of the BMWi through DLR grant 50 OR 0701.  The research in
this work has been partially funded by the European Commission under
grant ITN 215212.  S. Markoff acknowledges support from a Netherlands
Organization for Scientific Research (NWO) Vidi and OC Fellowship.
\end{acknowledgements}

\bibliographystyle{aa}

\begin{thebibliography}{}

\bibitem[\protect\astroncite{{Belloni} et~al.}{2005}]{belloni:05a}
{Belloni} T., {Homan} J., {Casella} P., et~al., 2005, A\&A 440, 207

\bibitem[\protect\astroncite{Beloborodov}{1999}]{beloborodov:99a}
Beloborodov A.M.,  1999, ApJ 510, L123

\bibitem[\protect\astroncite{Bowyer et~al.}{1965}]{bowyer:65a}
Bowyer S., Byram E.T., Chubb T.A., Friedman H.,  1965, Science 147, 394

\bibitem[\protect\astroncite{Coppi}{1999}]{coppi:99a}
Coppi P.,  1999,
\newblock PASP Conf. Series, 161, 375

\bibitem[\protect\astroncite{Dove et~al.}{1997}]{dove:97b}
Dove J.B., Wilms J., Maisack M.G., Begelman M.C.,  1997, ApJ 487, 759

\bibitem[\protect\astroncite{{Gou} et~al.}{2011}]{gou:11a}
{Gou} L., {McClintock} J.E., {Reid} M.J., et~al., 2011, ApJ submitted
  (arXiv:1106.3690)

\bibitem[\protect\astroncite{{Hanke} et~al.}{2009}]{hanke:08a}
{Hanke} M., {Wilms} J., {Nowak} M.A., et~al., 2009, ApJ 690, 330

\bibitem[\protect\astroncite{{Homan} et~al.}{2005}]{homan:05a}
{Homan} J., Buxton M., Markoff S., et~al., 2005, ApJ 624, 295

\bibitem[\protect\astroncite{{Ibragimov} et~al.}{2005}]{ibragimov:05a}
{Ibragimov} A., {Poutanen} J., {Gilfanov} M., et~al., 2005, MNRAS 362, 1435

\bibitem[\protect\astroncite{{Maitra} et~al.}{2009}]{maitra:09a}
{Maitra} D., {Markoff} S., {Brocksopp} C., et~al., 2009, MNRAS 398, 1638

\bibitem[\protect\astroncite{Markoff et~al.}{2005}]{markoff:05a}
Markoff S., Nowak M., Wilms J.,  2005, ApJ 635, 1203

\bibitem[\protect\astroncite{{Miller} et~al.}{2006}]{miller:06b}
{Miller} J.M., {Homan} J., {Steeghs} D., et~al., 2006, ApJ 653, 525

\bibitem[\protect\astroncite{Nowak et~al.}{2011}]{nowak:11a}
Nowak M.A., Hanke M., Trowbridge S.N., et~al., 2011, ApJ 728, 13

\bibitem[\protect\astroncite{Nowak et~al.}{2008}]{nowak:08a}
Nowak M.A., Juett A., Homan J., et~al., 2008, ApJ 689, 1199

\bibitem[\protect\astroncite{Nowak et~al.}{2005}]{nowak:05a}
Nowak M.A., Wilms J., Heinz S., et~al., 2005, ApJ 626, 1006

\bibitem[\protect\astroncite{{Orosz} et~al.}{2011}]{orosz:11a}
{Orosz} J., {McClintock} J., {Aufdenberg} J., et~al., 2011, ApJ submitted
  (arXiv:1106.3689)

\bibitem[\protect\astroncite{{Poutanen} \& Vurm}{2009}]{poutanen:09a}
{Poutanen} J., Vurm I.,  2009, ApJ 690, L97

\bibitem[\protect\astroncite{{Reid} et~al.}{2011}]{reid:11a}
{Reid} M.J., {McClintock} J.E., {Narayan} R., et~al., 2011 submitted
  (arXiv:1106.3688)

\bibitem[\protect\astroncite{Tananbaum et~al.}{1972}]{tananbaum:72a}
Tananbaum H., Gursky H., Kellogg E.M., et~al., 1972, ApJ 174, L143

\bibitem[\protect\astroncite{Wilms et~al.}{2006}]{wilms:06a}
Wilms J., Nowak M., Pottschmidt K., et~al., 2006, A\&A 447, 245

\bibitem[\protect\astroncite{Xiang et~al.}{2011}]{xiang:11a}
Xiang J., Lee J.C., Nowak M.A., Wilms J.,  2011, ApJ in press (arXiv:1106.3378)

\end{thebibliography}

\medskip
\noindent {\bf DISCUSSION}

\medskip
\noindent {\bf VALENT\'I BOSCH-RAMON:} You mentioned that
\cyg\ remains in the lower horizontal branch of the
hardness-luminosity q-diagram.  What about the blob detected in the
radio by Fender et al.? Wasn't it coming from a state transition in
the higher branch?

\medskip
\noindent {\bf MICHAEL NOWAK:} Whether or not ejection events
typically also accompany the state transition from spectrally
soft/radio quiet to spectrally hard/radio loud is still an open
question.  That being said, there's no evidence that \cyg\ has been on
the ``upper'' hard/soft transition branch (unless the two branches are
\emph{very} close to each other in terms of bolometric flux for \cyg).

\medskip
\noindent {\bf J\"ORN WILMS:} The flare was discussed by Wilms et
al. (2007, ApJ, 663, L97).  \cyg\ was close to a transition from its
soft state to its hard state, and the radio flare followed the X-ray
flare by $\approx 7$\,minutes.

\end{document}